\documentclass[%
 aip,
 amsmath,amssymb,
 reprint,%
]{revtex4-1}

\usepackage{graphicx}
\usepackage{dcolumn}
\usepackage{bm}
\usepackage{amsmath, amssymb, stmaryrd, xparse, mleftright}
\usepackage{color}
\usepackage{array}
\usepackage{ulem}
\usepackage[utf8]{inputenc}
\usepackage[T1]{fontenc}
\usepackage{mathptmx}
\usepackage{etoolbox}

\usepackage{blindtext, xcolor}
\usepackage{xcolor}
\usepackage{hyperref}
\hypersetup{
    colorlinks,%
    citecolor=blue,%
    linkcolor=blue,%
    urlcolor=blue
}

\begin{document}

\preprint{AIP/123-QED}

\title{RKKY interactions mediated by topological states in transition metal doped bismuthene}

\author{Emmanuel V. C. Lopes}
\email{emmanuelvictor96@gmail.com}
\affiliation{ 
Instituto de Física, Universidade Federal de Uberlândia, Uberlândia, Minas Gerais 38400-902, Brazil
}
\author{E. Vernek}%
\affiliation{Nanoscale and Quantum Phenomena Institute, and Department of Physics \& Astronomy, Ohio University, Athens, Ohio 45701, USA}
\affiliation{ 
Instituto de Física, Universidade Federal de Uberlândia, Uberlândia, Minas Gerais 38400-902, Brazil
}
\author{Tome M. Schmidt}
 \email{tschmidt@ufu.br}
\affiliation{ 
Instituto de Física, Universidade Federal de Uberlândia, Uberlândia, Minas Gerais 38400-902, Brazil
}

\date{\today}

\begin{abstract} 
We have investigated magnetic interactions between transition metal ions in bismuthene topological insulators with protected edge states. We find that these topological states have a crucial role in the magnetic interactions in 2D topological insulators. Using first-principles and model Hamiltonian, we make a comparative study of transition metal doped bulk and nanoribbon bismuthene. While a direct overlap between the transition metal prevails in gapped bulk bismuthene, at the borders of nanoribbons, a long-range magnetism is present. The exchange interactions are well described by a Ruderman-Kittel-Kasuya-Yosida-like Hamiltonian mediated by massive and topological states. Our results show a dominance of antiferromagnetism promoted by the topological states, preserving the spin-locked Dirac crossing states due to a global time-reversal symmetry preservation. This extended magnetic interactions mediated mainly by massless electrons can increase the spin diffusion length being promising for fast dissipationless spintronic devices.

\end{abstract}

\maketitle

\section{Introduction}
Since graphene experimental synthesis in 2004,\cite{Geim2007} the class of two-dimensional (2D) materials became an important subfield in condensed matter physics. A great deal of effort has been dedicated to predict and synthesize those materials within potential technology applications, such as quantum computing, spintronics, high speed and storage devices.\cite{Bader2010, Engel2001, Awschalom2007, Li2016} One special subclass of 2D materials is the quantum spin Hall (QSH) insulator, largely studied motivated by the presence of strong spin-orbit coupling (SOC), that leads to band inversion in bulk with spin-polarized edge channels protected by time-reversal symmetry. Since 2006, many of these new materials were predicted and synthesized. Examples encompass  HgTe, bismuthene, silicene, stanene\cite{Bernevig2006,Koenig2007,Murakami2006,Hirahara2011,Yang2012,Liu2011,Vogt2012,Zhu2015} and many others predicted, including  topological crystalline insulators.\cite{Hsieh2012,Chuang2013,Xu2013,Wrasse2014,Chuang2014,Hsu2016,Padilha2016}

Much of the interest in these 2D materials relies on the possibility of their application for practical purposes. Therefore, fine control of their intrinsic electronic and magnetic properties is pivotal. Fortunately, doping provides promising ways to manipulate electronic properties in 2D materials. More interestingly, it has been showed that magnetic properties can be modified by replacement of atoms in monolayers like graphene by cobalt or nitrogen,\cite{Gomes2012, Wang2011, Lv2012} for instance. As compared to normal 2D materials, QSH insulators
have an advantage for applications since massless spin-locked states are present at the edges or at the interfaces with trivial materials. As such, control over the spins of the conducting electrons has a consequence on the currents. Therefore, a detailed comprehension of interaction of electrons in these topological states with magnetic impurities is crucial to design devices for practical applications.

In this context, bismuthene is a promising QSH system, thanks to its large bandgap as compared to other 2D topological insulators (TIs).\cite{Wang2014,Lozovoy2022} Also the search for high speed dissipationless devices using bismuthene has become more intense after its synthesis on distinct surfaces using different techniques.\cite{Hirahara2011,Yang2012,Chuang2013,Sabater2013,Drozdov2014,Reis2017,Sun2021}
Moreover, bulk bismuthene doped with transition metals (TMs) is predicted to rich magnetic properties.\cite{Kadioglu2017,Qi2019,Muhammad2021,Saini2022} The TMs provide the bismuthene with magnetic impurities which, at a low density, can interact with each other leading to stable magnetic phases. What so far is poorly understood is the role played by topological states on the magnetic stability and, consequently, in the 1D spin current. The spin diffusion length must be greater than the device length scale to preserve the electron's spin.\cite{Hirohata2020} In this way, it is expected that long-range exchange interactions mediated by edge states may stabilize a magnetic order in the  system.\cite{Hermenau2019} Indeed, long-range Ruderman-Kittel-Kasuya-Yosida (RKKY) interactions\cite{Ruderman1954,Kasuya1956,Yosida1957} have been predicted in 3D topological crystalline insulator surfaces\cite{Yarmohammadi2020} and in doped MoS$_2$ flakes.\cite{Ulloa2016,Ulloa2018}

Motivated by the development progress of the 2D TI bismuthene, in this work, we used it as a platform to investigate the role played by topological states on the magnetic stability, by a direct substitution of bismuth atoms by TM ions. A comparative study of the magnetic correlations is performed in the bulk as well at the edges, where the topological phase has a direct impact on the exchange energy between impurities. With these results, we propose a low-energy effective Hamiltonian that accounts for the topological edge states and is coupling to the magnetic impurities. With this, upon integrating out the electrons' degrees of freedom, we derive an effective coupling between the impurities. The remarkable agreement between the RKKY-like couplings obtained from the effective model and the {\it ab initio} calculations  confirms  the role of  the topological edge states in the inter-impurity exchange interaction.

\section{Methodology}

We combine first-principles calculations and a model Hamiltonian to elucidate the effect of the topological states on the magnetic exchange interactions. The {\it ab initio} electronic structures were computed within density functional theory (DFT) using the projector augmented waves implemented in the VASP code.\cite{Kresse1993,Kresse1996} A cut-off energy of 420~eV of the kinetic energy for the plane wave basis set has been used. The exchange-correlation was described by the generalized gradient approximation within the Perdiew-Burke-Ernzehof's functional,\cite{Perdew1996} including fully relativistic pseudopotentials. To get a better description of the electronic correlation of the localized 3{\it d} TM orbitals, we use the DFT+U\cite{Anisimov1991} approximation. The onsite optimized effective Hubbard U parameter of 3.25~eV has been obtained for {\it d} orbitals.\cite{Dudarev1998, Cococcioni2005} The topological invariants have been computed using the Wannier charge center proposed by Soluyanov et al.\cite{Soluyanov2011, Gresch2017} The inter-impurity exchange interaction was studied by appropriate large unit cells, with up to 308 atoms in bulk and slab systems. The coupling between the TM impurities has been also interpreted by using model Hamiltonian\cite{Silva_2019} and compared with DFT results. These low-lying energy bands take into account the interaction between the TM impurities, including the topological edge states, through second order perturbation theory, as will be discussed in Sec.~\ref{effective hamiltonian}.

\section{Results and discussion}

We will use a vanadium ion as a prototype of 3{\it d} TM doped bismuthene to investigate the effects of the topological states on the magnetic interactions. The results can be extended to other 3{\it d} as well as to 4{\it d} TM ions. Vanadium doped bulk bismuthene forms covalent bonds with Bi host atoms, inducing a small inward relaxation of the Bi nearest neighbors. The bond-lengths of the Bi neighbors to the impurity is 2.81~{\rm\AA}, which is shorter than the pristine Bi-Bi distance of 3.12~{\rm\AA}, in agreement with previous results and similar to other TMs.\cite{Qi2019,Muhammad2021,Saini2022}

\begin{figure}[th]
    \centering
     \includegraphics[width=8.5cm, height =5.3cm]{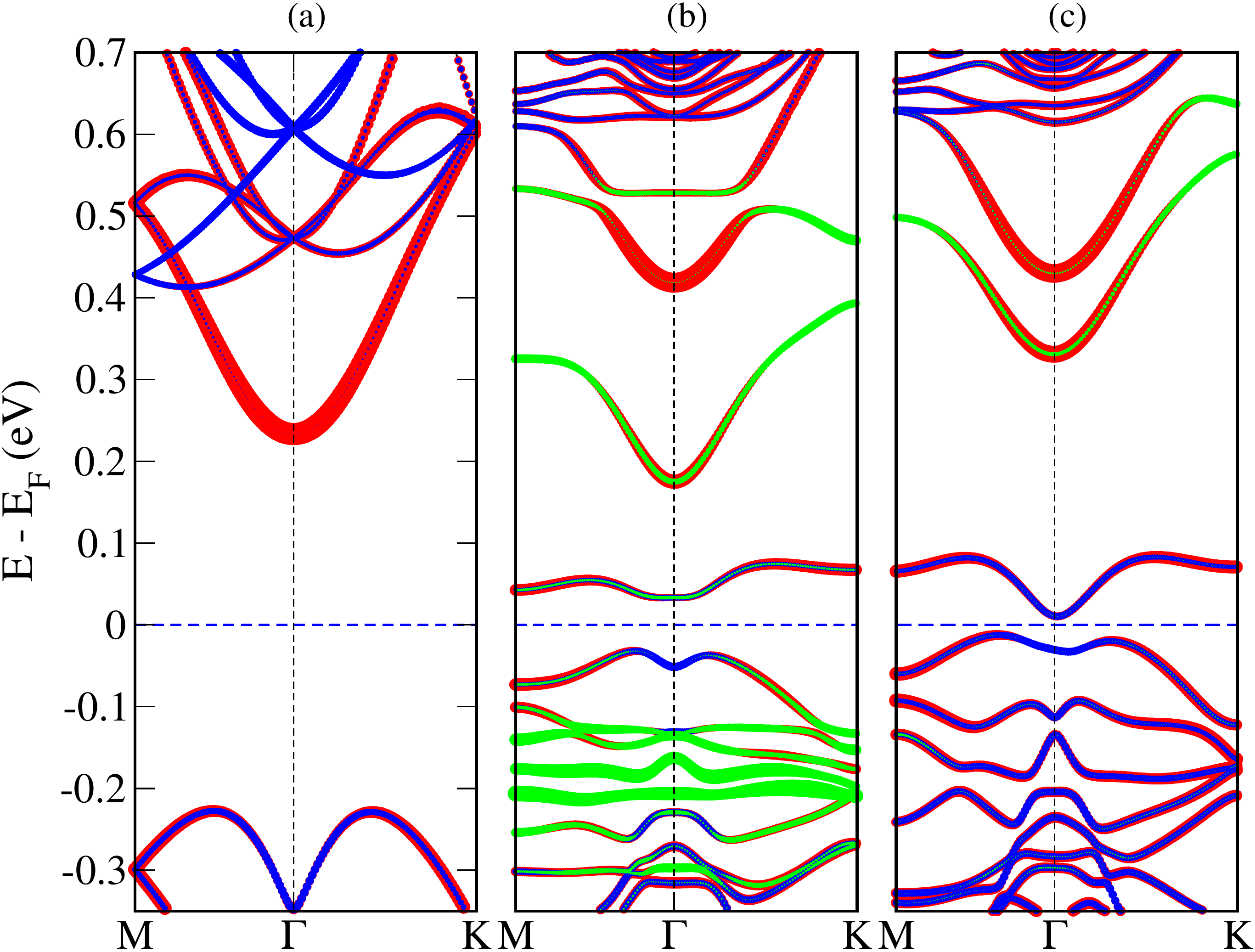}
      \caption{\label{Band}Band structure with atomic orbitals projected on $p_{xy}$ (red) and $p_z$ (blue) Bi ions, and transition metal $d$ (green) orbitals for: pristine bismuth monolayer (a), vanadium doped without (b) and with (c) effective U interaction. The Femi level (horizontal dashed line) is at zero energy.}
\end{figure}

Defects can modify the electronic properties of topological insulators.\cite{Kadioglu2017,Lima2016} As we can see in Fig.~\ref{Band}, a magnetic impurity induces spin splitting which is enhanced by the strong SOC interactions. The impurity introduces new energy levels inside the bulk bandgap, with strong {\it p-d} hybridization moving the Fermi level toward the valence band maximum (VBM) [Fig.~\ref{Band}(b)]. The correction in the Coulomb interactions by introducing the onsite effective U in the TM {\it d} orbitals opens up the bandgap, while keeping the {\it p}-type character [see Fig.~\ref{Band}(c)]. The topological non-trivial phase characterized by an inverted bandgap [showed for pristine bismuthene in Fig.~\ref{Band}(a)] is also better described with the introduction of the effective U interaction (compare Figs.~\ref{Band}(b) and \ref{Band}(c)). The topological gap is now between the impurity state and the split conduction band. The band structure shown in Fig.~\ref{Band} is for 2\% of V doping concentration. The topological gap is maintained by increasing the TM doping (checked up to 5\%), but it reduces the global bulk energy bandgap.\cite{Qi2019}

\subsection{Magnetic interactions mediated by TM doped gapped bulk bismuthene}

As the TM impurity introduces new energy levels around the Fermi level (as shown in Fig.~\ref{Band}), it brings a new possibility for magnetic research in bismuthene, once this substitution allows not only magnetic interaction between the ions, but also conduction electrons can maximize orbital overlap.\cite{Crook2018} Our calculations show that non-interacting TM ion doped bulk bismuthene has an out-of-plane magnetization of $3~\mu_B$ per ion. By increasing the number of impurities in an enlarged unit cell, we can investigate the magnetic impurity interactions. We first focus on the bulk doping, where V impurities have been diluted doped at Bi sites in a 2D periodic Bloch supercell. Figure~\ref{Total-energy}(a) shows the total energy difference between adjacent TM ions aligned ferromagnetically (FM) and antiferromagnetically (AFM) as a function of the distance between the impurities. The exchange interactions support a FM phase for most of the V-V distance in bulk bismuthene, with a small oscillation between the two phases. The most stable FM configuration occurs for a short-range exchange interaction for a V-V distance around 9~{\rm\AA} with the FM phase 120~meV more stable than the AFM one. This scenario is consistent with a predominant direct exchange interaction, however, the magnetic exchange energy is nonvanishing even for distances around 15 {\rm\AA} [Fig.~\ref{Total-energy}(a)], indicating some indirect magnetic exchange contribution that can be mediated by acceptor states [see Fig.~\ref{Band}(c)].

\subsection{Magnetic interactions mediated by topological states}

\begin{figure}
    \centering
    \includegraphics[width=8cm, height =9cm]{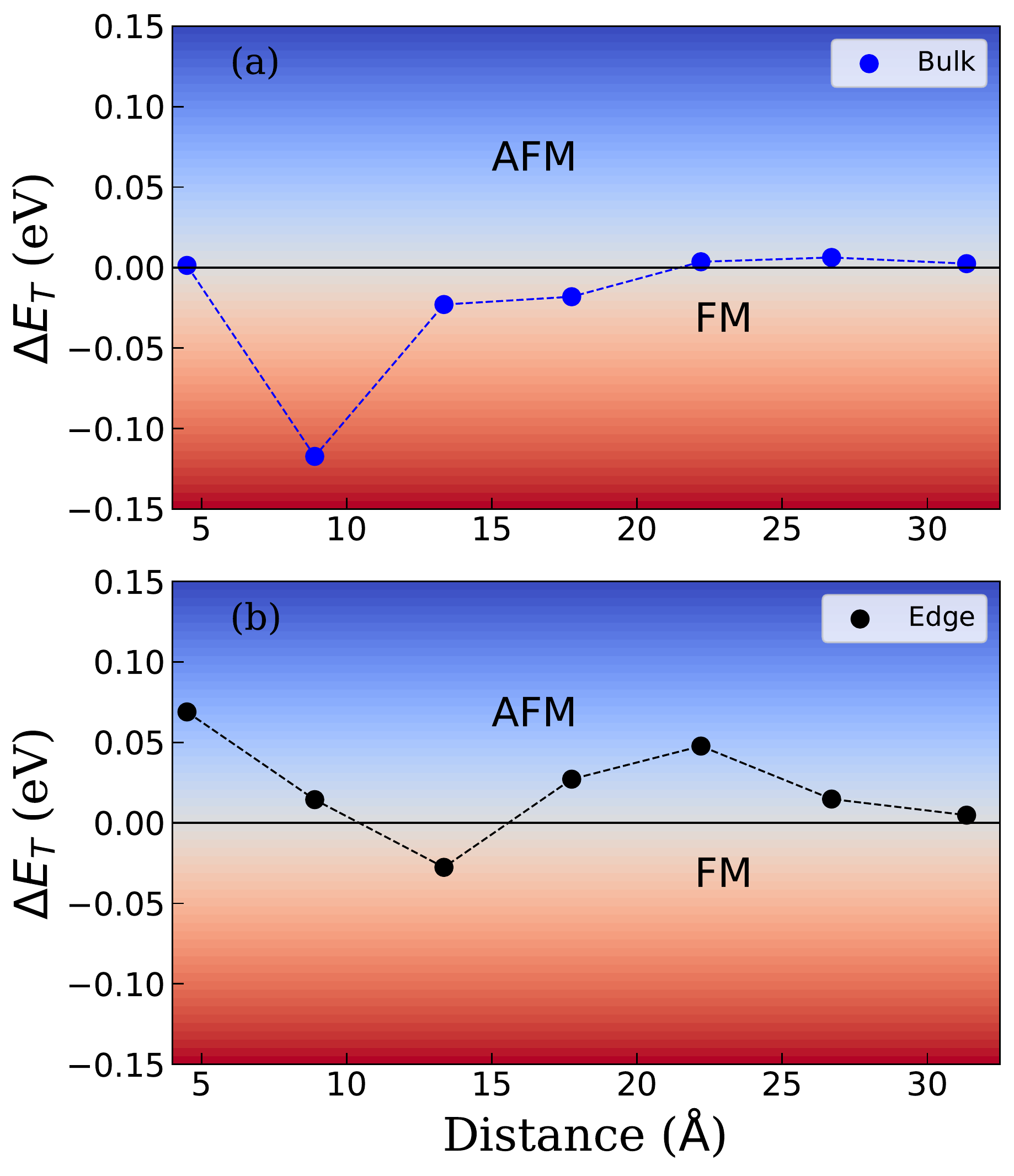}
     \caption{\label{Total-energy}Total energy difference ($\Delta E_T = E_T^{AFM}-E_T^{FM}$)  as function of the distance between TMs doped bulk (a) and edge (b) bismuthene.}
\end{figure}

The effect of massless spin-locked electrons on the magnetic interactions has been investigated by promoting TM doped at the borders of a bismuthene ribbon, where the topological edge states are. In contrast to TM doped bulk, the interaction between the impurities and topological states turns the magnetic phase strongly dependent on the TM distances, with a strong oscillation, as shown in Fig.~\ref{Total-energy}(b). The presence of topological states induces mostly antiferromagnetism, contrary to the gapped bulk system. The oscillation on the magnetic exchange interaction as a function of inter-impurity distance is one of the main characteristics of RKKY interaction, as predicted in doped bulk 2D materials.\cite{Crook2018,Crook2015scireports,Zare2016,Duan2018} 
The magnetic interactions in bismuthene shown in Fig.~\ref{Total-energy} show that the main contribution of the edge states is on the damping oscillation and not on the magnetic interactions magnitude. For the narrow bandgap silicene, a model Hamiltonian gave $10^5$ times larger RKKY coupling for magnetic impurities at the edge of the silicene as compared to the bulk.\cite{Zare2016} As we are using first-principles calculation here, we can conclude that a well defined bandgap reduces the RKKY strength, or the inclusion of all bands in a full relativistic self-consistent procedure is crucial to describe the effect of the edge states on the magnetic stability. DFT results for Fe-doped a 3D topological insulator Bi$_2$Se$_3$\cite{Kim2015} also show the same order of magnetic coupling in bulk and on the surface.
A refined description on the contribution of the topological states in the magnetic phase transitions is addressed in the next section by employing a RKKY-like Hamiltonian, whose parameters are taken from the first-principles results.

\begin{figure}
\centering
      \includegraphics[width=8.5cm, height=6cm]{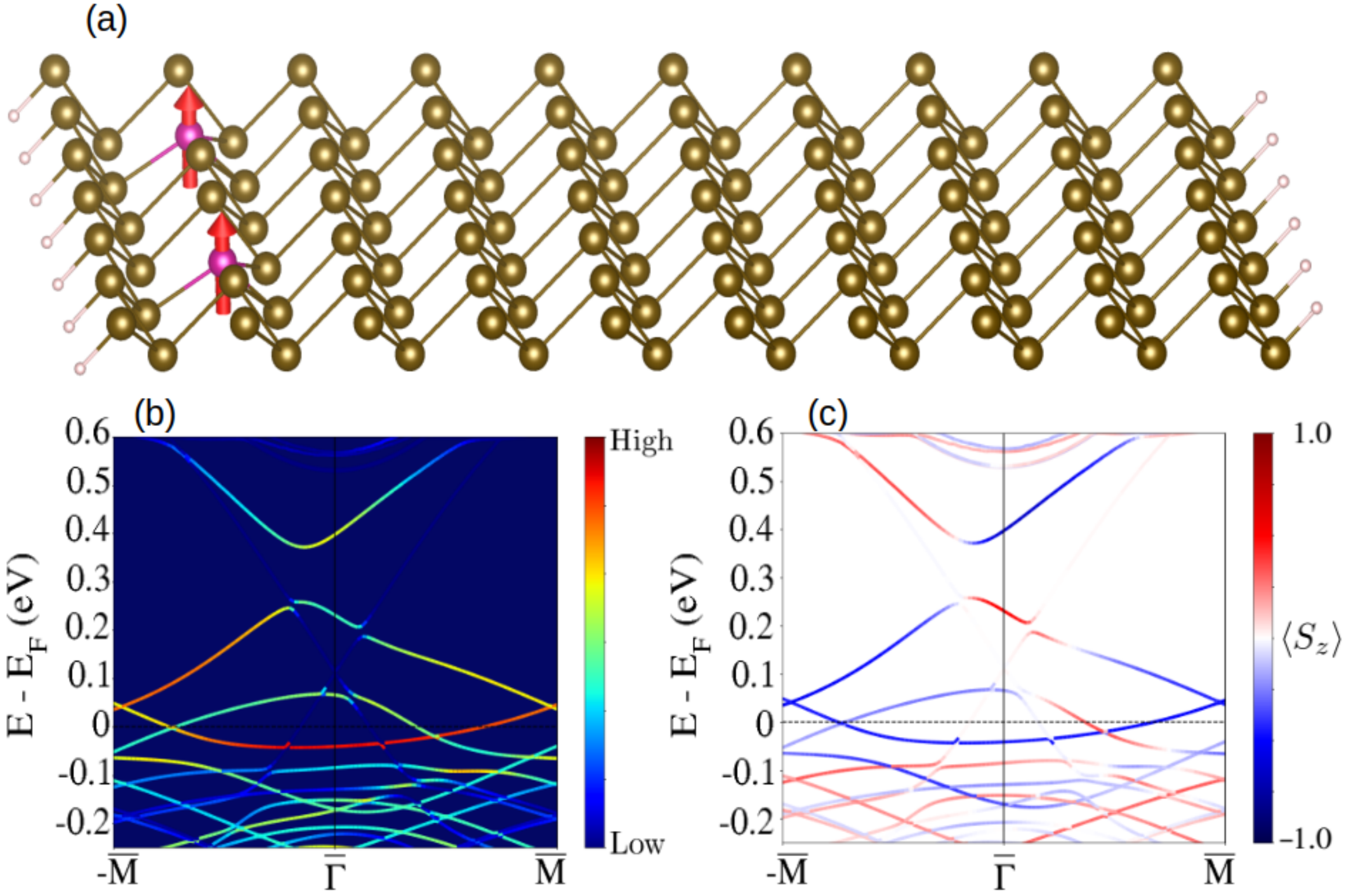}
      \includegraphics[width=8.5cm, height=6cm]{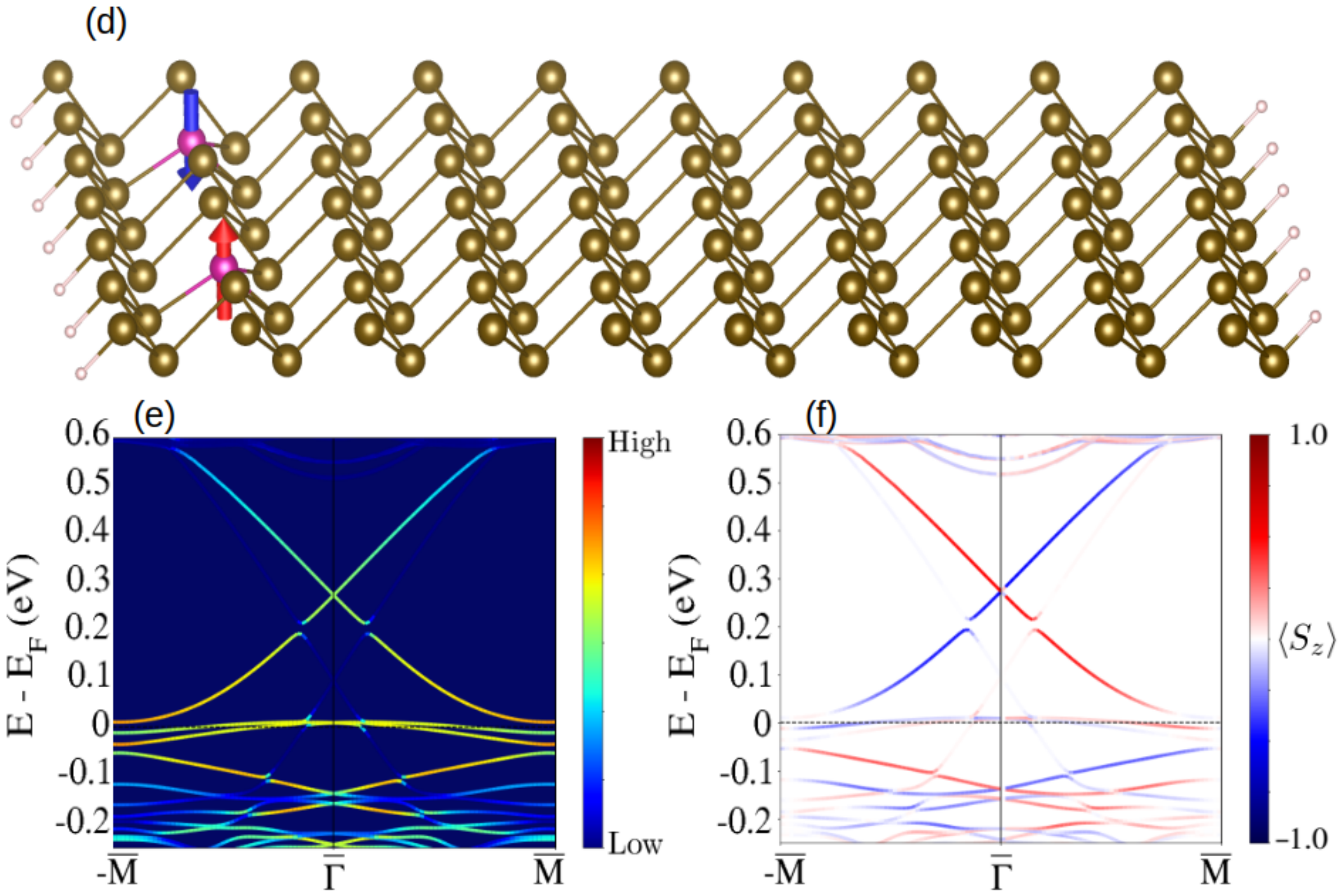}
    \caption{\label{spin-nanofita}Structural sketch of TM doped borders of a bismuthene nanoribbon with the FM phase (a). The nanoribbon has been saturated with H atoms to avoid edge dangling bonds and to preserve the Dirac crossing at the $\Gamma$ point.~\cite{Wang2014} Projection of border orbitals (b) and spin (c) for the FM phase. In (d)(f) are the same as above for TM aligned antiferromagnetically.}
    \label{fig:my_label}
\end{figure}

The effect of the RKKY interactions on the electronic properties has been investigated for two configurations of the magnetic ions, one FM and another AFM as a ground state. The FM coupling breaks time-reversal symmetry, opening up a gap in the topological Dirac states as can be seen in Fig.~\ref{spin-nanofita}(b). Although the spin-momentum is not locked anymore, there is a partial preservation of the topological state spin texture [see Fig.~\ref{spin-nanofita}(c)]. The easy axis of the impurity magnetization is out-of-plane, keeping the same magnetic moment as in the bulk, namely, $3~\mu_B$ per ion.
On the other side, impurities aligned antiferromagnetically [Fig.~\ref{spin-nanofita}(d)-\ref{spin-nanofita}(f)] keeps the Dirac crossing because of the global time-reversal symmetry preservation, $E\left(\textbf{k},\uparrow\right)=E\left(-\textbf{k},\downarrow\right)$. The AFM phase is more stable for most of inter-V distances [Fig.~\ref{Total-energy}(b)]. The surviving of edge Dirac states under the AFM phase was also obtained in Cr-doped bismuthene.\cite{Kim2018} The topological Dirac state spins are also aligned out-of-plane in this phase. The Fermi level is shifted downwards from the Dirac crossing point, where the bands now present linear and quadratic terms, similar to Rashba states.\cite{Bychkov1984} Both massless and massive electron states can contribute for the RKKY interactions. We compute the total energy without SOC (simulating a magnetic order without topological states at the edge) and the results are quite similar to the bulk ones, illustrating the importance of topological states on the long-range interactions.
This is indeed an important piece of information to propose the model that  will be discussed below.

\section{Effective Hamiltonian}\label{effective hamiltonian}

To further interpret our results obtained within DFT calculations, we derive an effective RKKY-like Hamiltonian that takes into account the shape of the low-lying energy bands of the system. To this end, we start by proposing a simple Hamiltonian $H_0$ that produces a linear dispersion of the pristine nanoribbon. Imagining the  simplest case where impurities will be mutually coupled indirectly via electrons that travels along the edge of the ribbon, such a Hamiltonian can be written as
\begin{equation}    
H_{0}=\sum_{k,s,s^\prime}(\beta k^2\delta_{ss'}-\alpha k\sigma_{ss'}^{z})c_{ks}^{\dagger}c_{ks'},   
\end{equation}
where $c_{ks}^{\dagger}$ ($c_{ks}$) represents the operator that creates (annihilates) an electron with spin $s$ in the lowest energy band and $\alpha$ is a parameter associated with the slope of the linear dispersion. The parameter $\beta$  is responsible for giving a small curvature to the band, since massive states are also present around the Fermi level from the DFT results. We now describe the coupling of the impurities with spin operators ${\bf S}_1$ and ${\bf S}_2$ to the electrons traveling along the edge as 
\begin{multline}
H_{j}=\frac{J}{N}\sum_{k,k'}e^{-i(k-k')x_{j}}[S_{j}^{z}\left(c_{k'\uparrow}^{\dagger}c_{k\uparrow}-c_{k'\downarrow}^{\dagger}c_{k\downarrow}\right)  \\
+S_{j}^{+}c_{k'\downarrow}^{\dagger}c_{k\uparrow}+S_{j}^{-}c_{k'\uparrow}^{\dagger}c_{k\downarrow}].
\end{multline}
Here, $x_j$ referring to the position of the  impurities along the edge of the ribbon, which we assume to lie along the $x$-direction. In addition, $S_j^z$ represents the $z$-component of the spin operator of the impurity $j$, while $S^\pm_j=\left(S^x_j\pm iS^y_j \right)/2$, in which $S^x_j$ and $S^y_j$ represent the spin operator for the $x$ and $y$ components, respectively. The total Hamiltonian is then $H=H_0+H_1+H_2$.

To obtain an effective Hamiltonian that describes the inter-impurity coupling mediated by the electrons moving along the edge of the ribbon, we employ a second order perturbation theory,\cite{Silva_2019} which encompasses integrating out the degrees of freedom of the itinerant electrons. Note here that the unperturbed Hamiltonian $H_0$ is already diagonal in the $\sigma^z$ basis. The resulting effective Hamiltonian can then be written as 
\begin{equation}
\label{H_eff}
\tilde H=I_{\parallel}S_{1}^{z}S_{2}^{z}+I_{\perp}S_{1}^{+}S_{2}^{-}+I^*_{\perp}S_{2}^{+}S_{1}^{-}.
\end{equation}
This corresponds to an anisotropic RKKY-like Hamiltonian where  $I_{\parallel}=2\mathrm{Re}\left(I_{++}+I_{--}\right)$ and  $I_{\perp}=\left(I_{-+}+I^*_{+-}\right)$, in which 
\begin{equation}\label{coupling}
I_{\delta\nu}=\frac{J^{2}}{4\pi^{2}}\int_{-k_\delta}^{k_\delta}dk\int_{\vert k'\vert>k_\delta}dk'\frac{e^{i(k-k')x}}{\beta(k^{2}-k'^{2})+\delta\alpha(k-\delta\nu k')}.
\end{equation}
In this expression $\delta, \nu \in \{+,-\}$, $k_\delta=k_F+\delta k_\alpha$, $k_\alpha=\alpha/2\beta$ and $x=x_2-x_1$ is the distance between the impurities, where we assume $x_2>x_1$ without loss of generality. 
Upon the convenient change of variable $k=qk_F$ and $q^\prime=k^\prime /k_F$, Eq.~\eqref{coupling} can be written in a better form
\begin{equation}\label{adm_coupling}
I_{\delta\nu}=I_{0}\int_{-q_{\delta}}^{q_{\delta}}dq\int_{\vert q'\vert>q_{\delta}}dq'\frac{e^{i(q-q')k_{F}x}}{(q^{2}-q'^{2})+\delta\tilde{a}(q-\delta\nu q')}.
\end{equation}
Here, $I_0=J^2/4\pi^2\beta$, $a=2k_\alpha/k_F=\alpha/\beta k_F$, and $q_\delta=1+\delta k_\alpha/k_F=1+\delta \tilde a/2$.

The integral in Eq.~\eqref{adm_coupling} is rather complicated but can be handled very analytically in very a formidable manner, as thoroughly discussed by one of us in Ref.~\onlinecite{Silva_2019}. Here, instead,  will perform the integration only numerically. To this end, we will choose the Hamiltonian parameters that best compare the result from the model with those from DFT calculations. In particular, a good choice is to set  $k_\alpha=0.05k_F$, which renders $\tilde a=0.1$. 

Our DFT calculations show that it is energetically favorable for the impurities' spins to align along the $z$-direction. Because of this, as mentioned above, for all calculations, the spins of the impurities were set along the $z$ axis. The indirect exchange interaction for this case is therefore only associated with the coupling $I_{\parallel}$ appearing in Eq.~\eqref{H_eff}.

\begin{figure}[t!]
    \centering
    \vskip0.5cm
    \includegraphics[width=8cm, height =5cm]{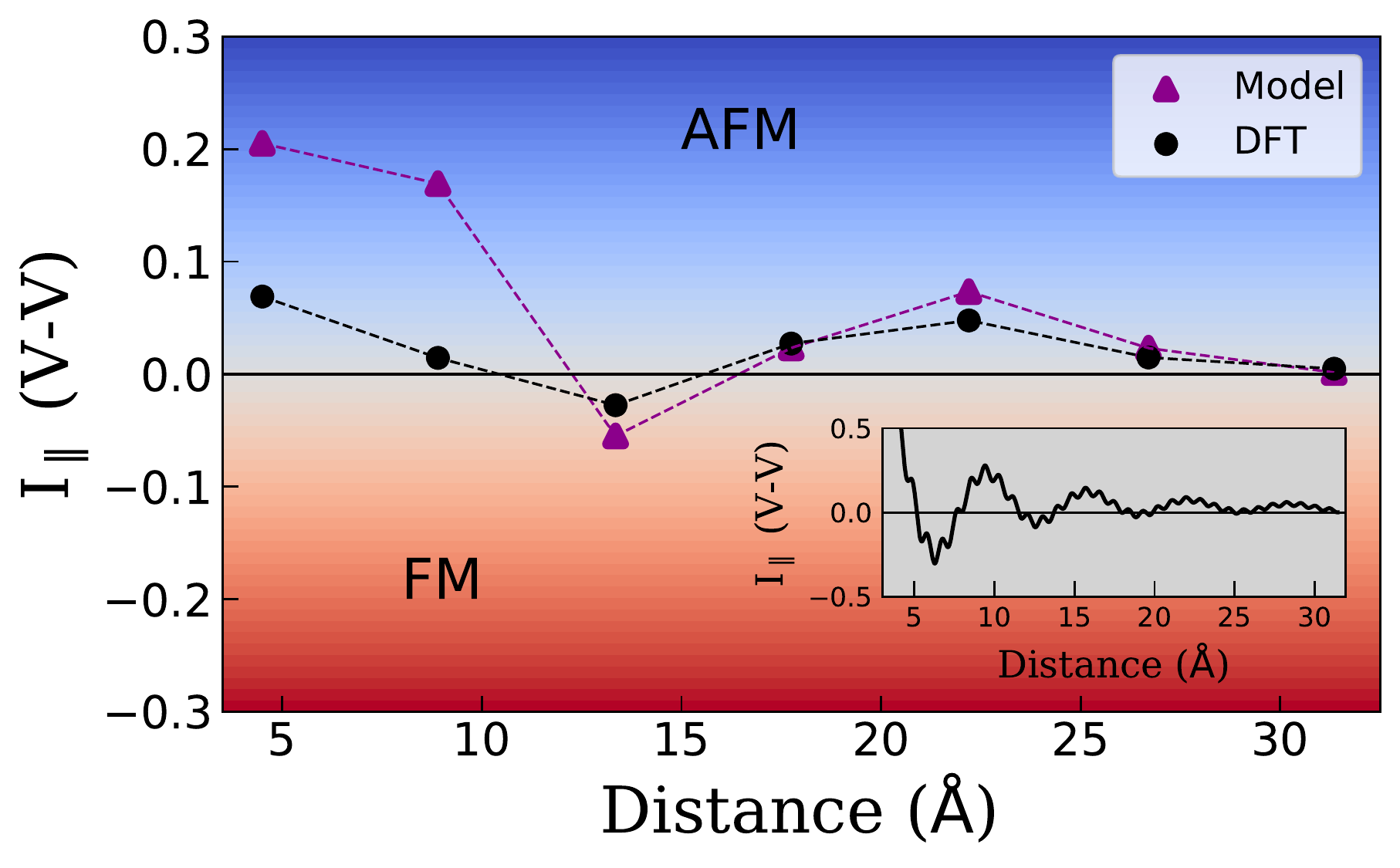}
     \caption{\label{RKKY-Effective}Direct parallel between DFT (black circles) and effective model (purple triangle) results for couple inter-V distances. Inset is the result of the model for a continuous range.}
\end{figure}

In Fig.~\ref{RKKY-Effective}, we depict the comparison between the energy $\Delta E_T$, obtained from DFT (black circles), and $I_{\parallel}$, calculated within the effective model (purple triangles), as a function of the distance between the impurities. Note that while the effective model provides results for any value of $x$ (the distance between the impurities), within the  DFT calculations, the positions should obey the crystal structure of the ribbon. We observe that the circles lie very close to the triangles, showing fairly good  agreement between the two approaches. Remarkably, even better agreement is observed for larger distances. The small deviation between the results for short distances may result from direct exchange interactions  provided by an overlap of the TM orbitals.  This good agreement suggests indeed that there is a RKKY-type of interaction between the impurities place close to the edge of the nanoribbon mediated mostly by topological states.

\section{Conclusion}

Based on first-principles and model Hamiltonian, we show the role played by topological states on the magnetic interactions in 2D topological insulators. Using TM doped bismuthene as the base for our investigation, we verify that doping gapped bulk and edge nanoribbons present distinct magnetic stability. Strong spin-orbit coupling usually reduces the spin diffusion length, which is the case in the bulk bismuthene. These short-range magnetic interactions are mostly promoted by a direct exchange coupling between the TM impurities. On the other side, by allowing interactions between the TMs and the topological states, more extended magnetic interactions are responsible for the magnetic stability. From our low-lying energy band model, the inter-impurity exchange interaction is well described by a RKKY-like Hamiltonian, mediated by a mixture of massive trivial and massless topological states. Our results also show a dominance of the AFM phase in the nanoribbon that preserves the spin-locked massless states, due to a global time-reversal symmetry preservation. Long-range magnetic interactions like RKKY extend the spin diffusion length making these systems potential for applications in high speed dissipationless devices, especially in spintronics.

\section{Acknowledgments}

The autors acknowledge Fundação de Amparo à Pesquisa do Estado de Minas Gerais, Coordenação de Aperfeiçoamento de Pessoal de Nível Superior, Conselho Nacional de Desenvolvimento Científico e Tecnológico and Instituto Nacional de Ciência e Tecnologia em Nanomateriais de Carbono for financial support, and also Laboratório Nacional de Computação Científica (LNCC-SCAFMat2) and Centro Nacional de Processamento de Alto Desempenho em São Paulo for the computational resources.

\bibliography{references.bib}


\end{document}